\documentclass[12pt,prl,preprint,letterpaper,linenumbers]{article}
\usepackage[english]{babel}
\usepackage[utf8]{inputenc}
\usepackage{lineno}
\usepackage{times}
\usepackage{amsmath}
\usepackage{titlesec}
\usepackage{fancyhdr}
\usepackage{subfig}
\usepackage{graphicx}
\usepackage{array}
\usepackage{tabularx}
\usepackage{adjustbox}
\usepackage{xspace}
\usepackage{xcolor}
\usepackage{authblk}
\usepackage{csquotes}
\usepackage{multirow}
\usepackage{amssymb}
\usepackage{physics}
\usepackage{accents}
\usepackage{float}
\usepackage{lipsum}
\usepackage[margin=1.0in]{geometry}
\usepackage[backend=biber,style=numeric,sorting=none,hyperref]{biblatex}
\title{Bubble Chamber Detectors with Light Nuclear Targets: A Snowmass 2021 White Paper}
\author[1]{Luis Alvarez-Ruso}
\author[2,3]{Joshua L. Barrow}
\author[4]{Leo Bellantoni}
\author[4]{Minerba Betancourt}
\author[4]{Alan Bross}
\author[5]{Linda Cremonesi}
\author[6,4]{Eric Dahl}
\author[7]{Kirsty Duffy}
\author[8]{Steven Dytman}
\author[9]{Laura Fields}
\author[10]{Tsutomu Fukuda}
\author[11]{Mikhail Gorchtein}
\author[12,4]{Richard J. Hill}
\author[4]{Alex Himmel} 
\author[4]{Thomas Junk}
\author[13]{Dustin Keller}
\author[14]{Huey-Wen Lin} 
\author[15]{Xianguo Lu} 
\author[14]{Kendall Mahn}
\author[16,17]{Aaron S. Meyer}
\author[4]{Jorge G. Morf\'{i}n}
\author[4]{Jonathan Paley}
\author[4,19]{Vishvas Pandey}
\author[20]{Gil Paz} 
\author[21]{Roberto Petti}
\author[12,4]{Ryan Plestid} 
\author[4]{Bryan Ramson}
\author[17]{Brooke Russell}
\author[22]{Federico Sanchez Nieto}
\author[12,4,23]{Oleksandr Tomalak}
\author[17]{Callum Wilkinson}
\author[24]{Clarence Wret}

\affil[1]{Instituto de F\'isica Corpuscular (IFIC)\\ Consejo Superior de Investigaciones Cient\'ificas (CSIC) and Universidad de Valencia (UV)\\ E-46980, Valencia, Spain}
\affil[2]{Massachusetts Institute of Technology, Cambridge, MA}
\affil[3]{Tel Aviv University, Tel Aviv, Israel}
\affil[4]{Fermi National Accelerator Laboratory, Batavia, IL 60510, USA}
\affil[5]{University College, London, London, WC1E 6BT, United Kingdom}
\affil[6]{Northwestern University, Evanston, IL 60208, USA}
\affil[7]{University of Oxford, Oxford, OX1 3RH, United Kingdom}
\affil[8]{University of Pittsburgh, Pittsburgh, PA 15260, USA}
\affil[9]{University of Notre Dame, Notre Dame, IN 46556, USA}
\affil[10]{IAR/Flab, Nagoya University, Furo-cho, Chikusa-ku, Nagoya, 464-8601, Japan}

\affil[11]{Universit\"at Mainz, 55122 Mainz, Germany}
\affil[12]{University of Kentucky, Department of Physics and Astronomy, Lexington, KY 40506, USA}
\affil[13]{University of Virginia, Charlottesville, VA 22904, USA}
\affil[14]{Michigan State University, East Lansing, MI 48824, USA}
\affil[15]{University of Warwick, Coventry, CV4 7AL, United Kingdom}
\affil[16]{University of California, Berkeley, CA 94720, USA}
\affil[17]{Lawrence Berkeley National Laboratory, Berkeley, CA 94720, USA}
\affil[19]{University of Florida, Gainesville, FL 32611-8440, USA}
\affil[20]{Wayne State University, Detroit, MI 48202 USA}
\affil[21]{University of South Carolina, Columbia, SC 29208, USA}
\affil[22]{Universit\'e de Gen\`eve, 1211 Geneva, Switzerland}
\affil[23]{Theoretical Division, Los Alamos National Laboratory, Los Alamos, NM 87545, USA}
\affil[24]{University of Rochester, Rochester, NY 14627, USA}

\begin{document}
\date{}
\maketitle

\begin{abstract}
Neutrino cross sections are a critical ingredient in experiments that depend on neutrino scattering to reconstruct event kinematics and infer neutrino characteristics, like NOvA and T2K.
An opportunity exists to reduce the 5-10\% broad uncertainty on neutrino cross sections by producing more measurements of neutrino scattering from light nuclear targets at the relevant energies. 
Bubble chambers with light nuclear targets would be ideal for these measurements but the most recent device designed for use with an accelerator neutrino source is at least fifty years old.
A new bubble chamber with light nuclear targets could be designed by observing how the technology has progressed for use in dark matter experiments and producing smaller modular devices that use more efficient cooling systems.
A smaller modular device could also be designed for deployment to all functioning neutrino beams, but an investigation of the proper operating characteristics is necessary to adapt newer detectors to the structure of contemporary neutrino beams.

\end{abstract}
\section{Introduction}
Most currently operating and proposed neutrino experiments indirectly rely on previously measured neutrino interaction cross sections to reconstruct the characteristics of events involving neutrinos.
For example, leading long-baseline neutrino oscillation experiments like T2K and NOvA directly use function extractions from world neutrino cross-section data in neutrino interaction generators to simulate event topology and detector response.% while conducting measurements of Pontecorvo-Maki-Nakagawa-Sakata (PMNS) matrix elements.
While the uncertainties on the current generation of long-baseline neutrino oscillation experiments are statistical, large uncertainties in neutrino-nucleus cross sections occupy a significant portion of the systematic error budgets on both experiments  \cite{aceroetal:2019hjg,abeetal:2019wsd}. 
Although manifesting in each experiment in different ways, the underlying problem is due to a broad uncertainty in neutrino cross-sections of between 5\% and 10\% \cite{alvarezrusoetal:2017qwr}.

%can be localized to the uncertainty inherent to neutrino scattering from many-nucleon targets.
%Uncertainties on cross sections must be reduced to 1\% or less for the next generation of experiments but this is unlikely to occur solely with measurements from the planned Near Detectors (ND) of DUNE or T2HK because neutrino scattering from the proposed nuclear targets of these detectors

Neutrino scattering from nuclei can be conceptually divided into two broad stages.
The first stage consists of the initial neutrino interaction with the nucleus, constituent nucleon(s), or quark(s) using one of multiple possible overlapping processes. 
The second stage is characterized by the secondary interactions of the products of the first stage with cold nuclear matter as they propagate through the nucleus (or nucleon)--dubbed Final State Interactions (FSI) \cite{formaggio:2012xer}. 
The overwhelming majority of existing world-data was collected with relatively heavy nuclei, embedding significant model dependence in cross-section extractions and reducing the exclusionary power of generators using these cross sections.
Also, in many cases, direct measurements on light nuclear targets serve as direct inputs into analyses. 
A significant reduction in uncertainties can come from making new dedicated measurements of neutrino scattering on light nuclear targets like hydrogen and deuterium \cite{alvarezrusoetal:2022wdv}. 
At a minimum, the first and second stages of interaction in neutrino scattering can be factorized.

The most recent detectors designed to use light nuclear targets are 50-year-old hydrogen bubble chamber experiments like the Fermilab 15' Bubble Chamber, the Argonne National Laboratory 12' Bubble Chamber, and the Brookhaven National Laboratory 80" Bubble Chamber\cite{binghametal:1990bkl, jaeger:1974try, jensen:1964cvn}.
The latest results from these experiments are an important (albeit small) contribution to world data, but the field has significantly progressed since these data were collected and analyzed.
Even modern reanalysis designed to improve the precision of bubble chamber data shows significant tension, negatively contributing to the overall uncertainty of the cross-section extractions supporting event generators \cite{wilkinsonetal:2014scv}.

It is possible that a new bubble chamber project could provide a complementary neutrino scattering dataset to the Near Detector (ND) programs of both DUNE and T2HK, significantly increasing the precision of their flagship measurements through the deconvolution of initial and final state neutrino-nucleus interactions \cite{abietal:2020olk, abeetal:2018ret}.
If used with the ultra-high intensity beams expected in the next generation of precision neutrino experiments, a new bubble chamber could make large contributions to nuclear physics, including broadly contributing to an understanding of nuclear structure however historic designs would miss significant portions of beam due to the cooling required for continuous cycling. 
This paper proposes the development of a new device which will use many of the design principles of historic devices along with ideas from contemporary dark matter focused chambers at a much smaller size to reduce cost and bulk cooling requirements.
Critical focus for this project consists of finding the best operating characteristics of a new modular device for application in a wide selection of neutrino beams.

\section{Bubble Chamber Operation and Characteristics}
\begin{figure}[htp!]
    \centering
    \includegraphics[width=0.48\textwidth]{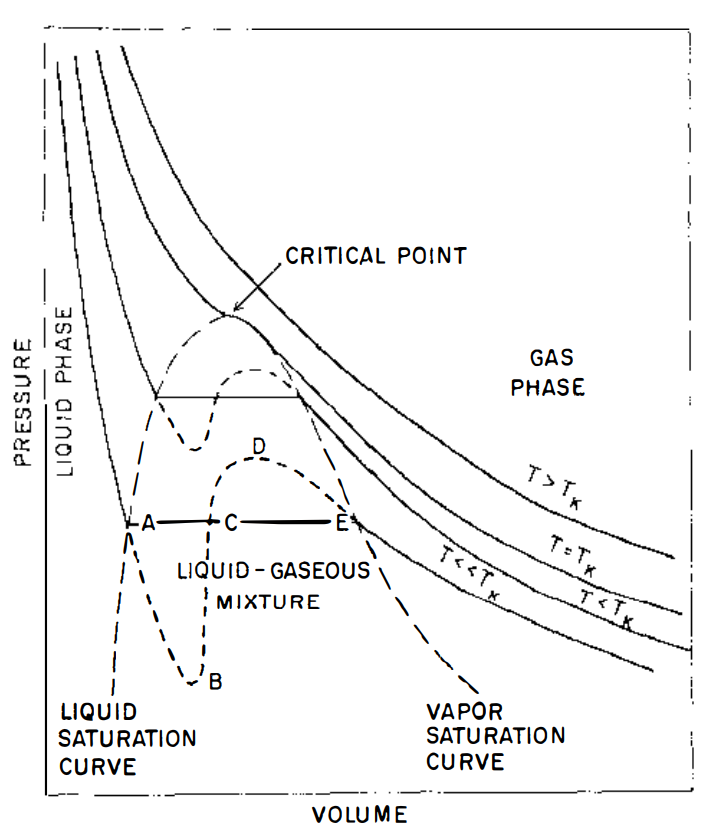}
    \caption{\textbf{A Brief Review of Bubble Chamber Thermodynamics} (From Ref.~\cite{bradner:1960rkl}) Ideally, a bubble chamber would start operation with a working fluid in the liquid phase well below the critical temperature, T$_{k}$. In a relatively uncontrolled environment with similar fluid and conditions (equivalent temperature and pressure, and with T$<<$T$_{k}$), as volume expanded, a boiling fluid would randomly see bubble nucleation through any imperfection at the fluid boundary. At this stage any changes in volume produce more gas until the entire liquid had been boiled off at point E. In a bubble chamber with a smooth boundary, the working fluid would follow paths A to B, changing phase to point D only as a consequence of energy deposition from charged particles. The energy required to "activate" the medium is directly proportional to the Gibbs Free Energy of the system.}
    \label{fig:thermo}
\end{figure}

Bubble chambers operate through the manipulation of a working fluid that begins in a superheated state at a preferred temperature through pressurization with a movable piston. 
Before observing events, the working fluid is shifted into a meta-stable state through piston movement to a higher volume, significantly reducing the pressure on the working fluid and allowing the device to occupy a sensitive state where energy deposition into the medium above a certain threshold produces bubbles \cite{glaser:1952der, seitz:1958jmp}. 
A representation of the state of the working fluid through the operation of the bubble chamber is reproduced from a cited reference in Figure \ref{fig:thermo} \cite{bradner:1960rkl}.
After the appearance of bubbles, which traced particle tracks, a picture was taken by a traditional camera.

\begin{figure}[htp!]
    \centering
    \includegraphics[width=0.45\textwidth]{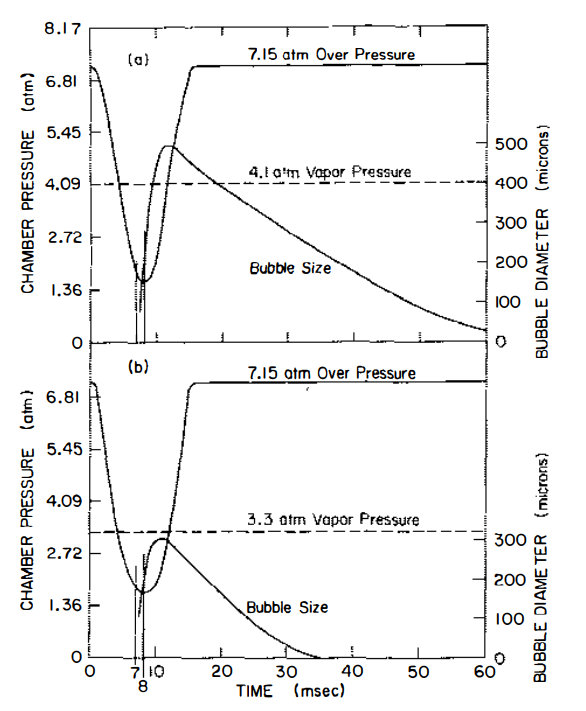}
    \caption{\textbf{Bubble Chamber Sensitivity and Bubble Size} (From Ref.~\cite{ballam:1977qrt}) The figure shows bubble chamber sensitivity and bubble survival time as a function of bubble size for the SLAC 1m Hydrogen Hybrid Bubble Chamber. The working fluid is protium. The top plot shows piston expansion (a pressure drop) with a maximum allowed bubble size of 450~$\mu$m and the bottom plot shows the same sensitivity but with a maximum allowed bubble size of 300~$\mu$m. While the over pressure in both instances remains the same, changing the operating state of the working fluid allows for quicker cycling. }
    \label{fig:bubblesize}
\end{figure}

Device operation is heavily dependent on the working fluid and its operating temperature which had a direct effect on the Gibbs free energy of the system. 
Both properties (the temperature and Gibbs free energy) are directly related to the energy deposition threshold for bubble nucleation. 
An examination of bubble size and detector sensitivity for the SLAC 1m Hybrid Chamber, which was partially filled with hydrogen, is reproduced from a cited reference in \ref{fig:bubblesize} \cite{ballam:1977qrt}.
Offline analysis was typically done by human observers who identified relevant events \emph{by eye}, and processed analysis with the earliest computers.
Raw images from bubble chambers were exceptionally precise and, with meticulous reconstruction of events, yielded high-precision event-by-event reconstruction.
The standard for spatial and momentum resolution in historic devices were less than a millimeter and around 2\% of the total momentum, respectively \cite{bradner:1960rkl}.
Specialized cameras and holographic techniques could increase the spatial resolution to less than 10~$\mu$m \cite{binghametal:1990bkl}.

\begin{figure}[htp!]
    \centering
    \includegraphics[width=0.75\textwidth]{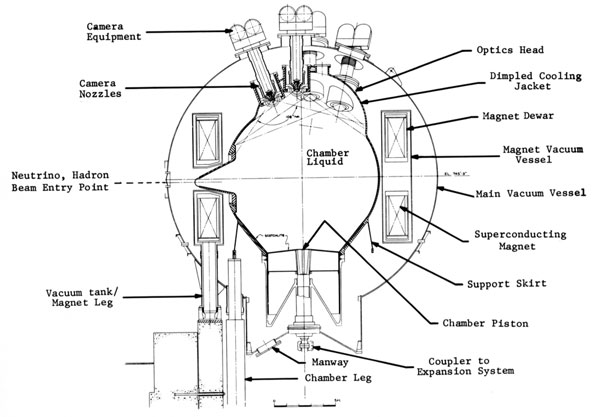}
    \caption{\textbf{Schematic of the 15' Bubble Chamber} (From Ref.~\cite{binghametal:1990bkl})The figure shows a schematic of the Fermilab 15' Bubble Chamber and typifies the design of dirty bubble chambers. This bubble chamber was last of the large bubble chambers and concluded operation on 1988. In this design, the primary working fluid was hydrogen and deuterium.}
    \label{fig:15footer}
\end{figure}

The most recent accelerator focused design was the hydrogen filled Fermilab 15' Bubble Chamber, which observed neutrinos generated by the historic 400 GeV Fermilab proton synchrotron.
The Fermilab 15' Bubble Chamber was characterized by a simplified design with stainless-steel inner chambers and synced to a known beam arrival time.
The chamber was rapidly cycled to prevent boiling and could sustain a 1 Hz cycling rate for 10 seconds before requiring a 50 second downtime to reset however, the device was mechanically capable of cycling at a rate of 3 Hz.
The limiting factor for a continuously cycling configuration was the cooling capacity of the chamber which was 6.6~kW \cite{shutt:1967sdf,baltay:1976mst}. 
A schematic of the Fermilab 15' Bubble Chamber is shown in Figure \ref{fig:15footer}.

\begin{figure}[htp!]
    \centering
    \includegraphics[width=0.36\textwidth]{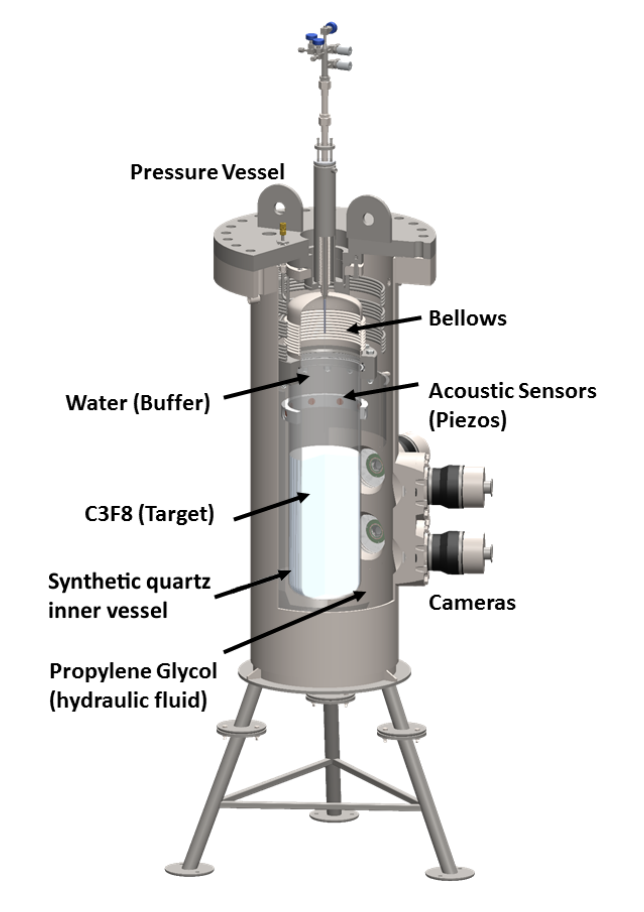}
    \caption{\textbf{Schematic of the PICO-60 bubble chamber} (From Ref.~\cite{amoleetal:2019oil}) The figure shows a schematic of the PICO-60 bubble chamber. This chamber design is one of the new generation of clean bubble chambers designed to detect dark matter. Newer chambers use silica glass to contain the target fluid, which in this case is C$_{3}$F$_{8}$. Water is used as a buffer fluid to manipulate the state of the target fluid. This particular chamber can be triggered acoustically. }
    \label{fig:pico60l}
\end{figure}

Recent prototype dark matter-focused bubble chambers constructed by the PICO and Scintillating Bubble Chamber (SBC) collaborations have unequivocally demonstrated that advances in sensor technology, computing, and automation can match or exceed the precision of historic chambers but newer chamber designs operate with a key difference compared to historic devices.
Newer chambers are all “clean” style devices, deploying silica or glass inner vessels with microscopically “smooth” contact surfaces to reduce the number of bubble nucleation sites (sites for uncontrolled boiling) at the edges of the detector \cite{bressleretal:2019ghr, baxteretal:2017plq, flores:2021zny}. 
A schematic of the PICO-60 prototype is reproduced from a cited reference in Figure \ref{fig:pico60l} \cite{amoleetal:2019oil}.

This design feature allows the superheated sensing medium to remain active for longer periods of time (minute to hours) at higher pressures and relative Gibbs free energy compared to historic accelerator focused, "dirty", bubble chambers \cite{bugg:1959tre}. 
Current clean chambers are unsuitable for the observation of neutrino scattering--lower pressures, higher temperatures, and a completely different range of Gibbs free energies allows for long sensitivity times but take much longer to cycle than historic accelerator focused "dirty" chambers.

\section{A Modern Bubble Chamber Concept}

A new chamber must be designed for the observation of neutrinos from modern source of neutrinos.
Many important questions arise concerning the m  ost cost-effective way of constructing a new device or \emph{devices}, and once a functioning device exists, where to put it. 
One idea is the development of a modular device of a standard size (around 1 ton of protium with a 1.5 m characteristic length), with embedded cryogenic support, piston, magnet, and edge detection through embedded scintillator. 
A single device could be deployed as a beam target for a larger suite of detectors or arrays of such a device could be deployed as a single detector (a la ArgonCube), improving the overall number of events to be collected \cite{majumdar:2021xse}. 
Ideally, a new hydrogen bubble chamber would be placed in beam from the Long Baseline Neutrino Facility (LBNF) but many options for accelerator based neutrino sources exist including the Neutrinos at the Main Injector (NuMI) beam, the Booster Neutrino Beam (BNB), and the J-PARC Main Ring. 

While the beam design has not yet been finalized, its spill structure can be approximated considering the spill structure of the NuMI beam \cite{adamsonetal:2016nbf}. 
The NuMI beam has an approximate spill length of ten microseconds and repetition rate on the scale of a single Hertz.  
A minimally sufficient new device would be a simple dirty style chamber with similar operating regime to the older chambers (4-10 atmospheres at around 20K) that could indefinitely cycle at 1 Hz and would be sensitive for at least 10 milliseconds per expansion.
An array of ten or more of these devices could yield millions of events over the lifetime of the DUNE/LBNF program and create a neutrino sample with complementarity to work on the upcoming Electron Ion Collider (EIC), reintroducing it as a novel probe of nucleon structure because of its clean access to axial vector current observables \cite{abdulkhaleketal:2021lpe}.

The Fermilab BNB would require fast cycling (greater than 10 Hz) or devices with longer sensitivity times and novel triggering techniques \cite{aguilararevaloetal:2009ojw}. 
BNB has a shorter spill length than the NuMI beam (less than 2 microseconds) but a repetition rate between 5 and 15 Hz. 
While this cycling time was mechanically possible for smaller historic dirty chambers, this choice would dramatically increase the required cooling load. 
A much simpler clean style device would have longer sensitivity and be active for every neutrino (as opposed to every beam spill) but would require an interior coated with Teflon, precise electropolishing, or metallic electroplating. 
It would also require the development of an operating thermodynamic regime somewhere between the preferred states of the historic dirty chambers and current clean chambers.
A Fermilab Lab Directed Research \& Development (LDRD) project investigating the best design for a bubble chamber with light nuclear targets has been funded and will investigate the feasibility of adapting different operating states for different beam lines.

\section{Conclusion}
Uncertainty on neutrino-nucleus cross sections affects all experiments that depend on neutrino scattering to infer neutrino properties, including all current and proposed long-baseline neutrino oscillation experiments.
Broad uncertainties on neutrino-nucleus cross sections are between 5\% and 10\% but can be greatly improved through measurements of neutrino scattering on light nuclear targets like hydrogen and deuterium.
Historic examples of devices with light nuclear targets include bubble chambers produced and hosted by Fermilab, Argonne, and Brookhaven National Laboratories. 
The most recent of these examples, the Fermilab 15' Bubble Chamber was a dirty bubble chamber, characterized by cycling that was timed to beam structure.
The Fermilab 15' Bubble Chamber had short sensitivity times, relatively short cycling times, and a long reset time relative to the beam due to cooling limitations.

A new generation of clean bubble chambers like the PICO-60 prototype offer an example of how to approach a contemporary bubble chamber with light nuclear targets but have heavy nuclear targets and are not designed to cycle on the timescales associated with accelerator based neutrinos sources.
A new design would use the progress made from contemporary devices to adapt historic dirty bubble chambers to existing and future beamlines. 
Neutrino beams that could host a bubble chamber currently exist at Fermilab (NuMI Beam and BNB) and J-PARC but the option that maximizes the statistics of a new experiment would be a new dedicated hall in the DUNE/LBNF beam.
Important bubble chamber design characteristics for accelerator based neutrino sources are modular devices with either long sensitivity times enabled by smooth chamber walls and novel triggering or quick cycling times enabled by efficient cooling and favorable states of the working fluid.
A Fermilab LDRD has been funded which will investigate the feasibility of developing a modern modular bubble chamber for light nuclear targets.

\printbibliography[title=References]

@article{aceroetal:2019hjg,
      author         = "Acero M. A. \emph{et al.} (NOvA Collaboration)",
      title          = "{First Measurement of Neutrino Oscillation Parameters using Neutrino and Antineutrinos by NOvA}",
      journal        = "Physical Review Letters",
      year           = "2019",
      volume         = "123",
      issue          = 15
}

@article{abeetal:2018ret,
      author         = "K. Abe \emph{et. al.}",
      title          = "{Hyper-Kamiokande Design Report}",
      journal        = "arXiv:1805.04163v2",
      year           = "2018",
      archivePrefix  = "arXiv:1805.04163v2"
}

@article{abietal:2020olk,
      author         = "Abi, B. \emph{et. al.} (DUNE Collaboration)",
      title          = "{DUNE Far Detector Technical Design Report, Volumes I, III, and IV}",
      collaboration  = "DUNE",
      journal        = "JINST",
      year           = "2020",
      volume         = "15",
      issue          = "08",
      page           = "T08008, T08009, T08010",
      %eprint         = "",
      archivePrefix  = "arXiv:2002.03010v1",                                                    
      primaryClass   = "physics.ins-det",
      reportNumber   = "FERMILAB-PUB-20-024-ND, FERMILAB-PUB-20-025-ND, FERMILAB-PUB-20-026-ND, FERMILAB-PUB-20-027-ND ",
      SLACcitation   = "%%CITATION = ARXIV:1706.09990;%%"
}

@article{alvarezrusoetal:2017qwr,
      author         = "L. Alvarez-Ruso \emph{et al.}",
      title          = "{NuSTEC White Paper: Status and Challenges of Neutrino-Nucleus Scattering}",
      journal        = "arXiv:1706.03621v2",
      year           = "2017",
      %eprint         = "",
      archivePrefix  = "arxXiv:1706.03621v2"                                                    
    
}

@article{wilkinsonetal:2014scv,
      author         = "C. Wilkinson \emph{et. al.}",
      title          = "Reanalysis of bubble chamber measurements of muon-neutrino induced single pion production",
      journal        = "Physical Review D",
      year           = "2014",
      issue          = "90",
      page           = "112017"
}

@article{formaggio:2012xer,
      author         = "G. P. Zeller, J. A. Formaggio",
      title          = "{From eV to EeV: Neutrino Cross Sections Across Energy Scales}",
      journal        = "Reviews of Modern Physics",
      year           = "2012",
      volume         = "84",
      issue          = "3",
      page           = "1307"
}

@article{alvarezrusoetal:2022wdv,
      author         = "L. Alvarez-Ruso \emph{et al.}",
      title          = "{Neutrino Scattering Measurements on Hydrogen and Deuterium: A Snowmass White Paper}",
      year           = "2022"
}

@article{binghametal:1990bkl,
      author         = "H. Bingham \emph{et al.}",
      title          = "{Holography of particle tracks in the Fermilab 15-Foot Bubble Chamber}",
      journal        = "Nuclear Instruments and Methods in Physics Research",
      year           = "1990",
      volume         = "A297",
      page           = "364-389"
}

@article{jaeger:1974try,
      author         = "K. Jaeger",
      title          = "{Performance of the Argonne 12-Foot Bubble Chamber}",
      journal        = "Conference Proceedings Track Sensitive Target Construction and Operating Conditions",
      year           = "1974"
}

@article{jensen:1964cvn,
      author         = "J. E. Jensen",
      title          = "{The Brookhaven National Laboratory 80-Inch Liquid Hydrogen Bubble Chamber}",
      journal        = "Cryogenic Engineering Conference No. E-2",
      year           = "1964"
}

@article{glaser:1952der,
      author         = "D. A. Glaser",
      title          = "{Some Effects of Ionizing Radiation on the Formation of Bubbles in Liquids}",
      journal        = "Physical Review",
      year           = "1952",
      volume         = "87",
      issue          = "4",
      page           = "665",
      eprint         = "https://doi.org/10.1103/PhysRev.87.665"   
}

@article{seitz:1958jmp,
      author         = "F. Seitz",
      title          = "{On the Theory of the Bubble Chamber}",
      journal        = "The Physics of Fluids",
      year           = "1958",
      volume         = "1",
      issue          = "1"
}

@article{bradner:1960rkl,
      author         = "H. Bradner",
      title          = "{Bubble Chambers}",
      journal        = "Annual Review of Nuclear Science",
      year           = "1960",
      volume         = "10",
      issue          = "109-160",
      eprint         = "https://doi.org/10.1146/annurev.ns.10.120160.000545"
}

@article{ballam:1977qrt,
      author         = "J. Ballam and R. D. Watt",
      title          = "{Hybrid Bubble Chamber Systems}",
      journal        = "Annual Review of Nuclear Science",
      year           = "1977",
      volume         = "27",
      issue          = "75-138",
      eprint         = "https://doi.org/10.1146/annurev.ns.27.1201177.000451"
}

@article{shutt:1967sdf,
      author         = "R. P. Shutt",
      title          = "",
      journal        = "{Bubble and Spark Chambers, Principles and Use, Vol. 1 \& 2}",
      year           = "1967",
}

@article{abeetal:2019wsd,
      author         = "K. Abe \emph{et al.}",
      title          = "First Measurement of Neutrino Oscillation Parameters using Neutrinos and Antineutrinos by NOvA",
      journal        = "Physics Review Letters",
      year           = "2019",
      volume         = "123",
      issue          = "15"
}

@article{baltay:1976mst,
      author         = "C. Baltay; W. B. Fowler; and D. Theriot",
      title          = "{Optimizing the Fermilab 15' Bubble Chamber}",
      journal        = "Proceedings of the NAL Summer Study on Utlization of the Energy Doubler/Saver",
      year           = "1976",
      volume         = "2"
}

@article{bressleretal:2019ghr,
      author         = "M. Bressler \emph{et al.}",
      title          = "A buffer-free concept bubble chamber for PICK dark matter searches",
      year           = "2019",
      eprint         = "arXiv:1905.07367v3"
}

@article{baxteretal:2017plq,
      author         = "D. Baxter \emph{et al.}",
      title          = "{First Demonstration of a Scintillating Xenon Bubble Chamber for Detecting Dark Matter and Coherent Elastic Neutrino-Nucleus Scattering}",
      year           = "2017",
      eprint         = "arXiv:1702.08861v2"
}

@article{flores:2021zny,
      author         = "L. J. Flores",
      title          = "{Physics reach of a low threshold scintillating argon bubble chamber in coherent elastic neutrino-nucleus scattering reactor experiments}",
      year           = "2021",
      eprint         = "arXiv:2101.08785v2"
}

@article{amoleetal:2019oil,
      author         = "C. Amole \emph{et al.}",
      title          = "{Dark matter search results from the complete exposure of the PICO-60 C$_{3}$F$_{8}$ bubble chamber}",
      journal        = "Physical Review D",
      year           = "2019",
      volume         = "100",
      issue          = "0220001",
      eprint         = "https://doi.org/10.1103/PhysRevD.100.022001"
}

@article{bugg:1959tre,
    author           = "D. V. Bugg",
      title          = "{The Bubble Chamber}",
      journal        = "Progress In Nuclear Physics",
      year           = "1959",
      volume         = "7",
      issue          = "1-52"
      %eprint         = "https://doi.org/10.1103/PhysRevD.100.022001"
}

@article{adamsonetal:2016nbf,
      author         = "P. Adamson \emph{et al.}",
      title          = "{The NuMI Neutrino Beam}",
      journal        = "Nuclear Instruments and Methods in Physics Research Section A: Accelerators, Spectrometers, Detectors, and Associated Equipment",
      year           = "2016",
      volume         = "806",
      page           = "279"
      
      %eprint         = "https://doi.org/10.1103/PhysRevD.100.022001"
}

@article{aguilararevaloetal:2009ojw,
      author         = "A. A. Aguilar-Arevalo \emph{et al.}",
      title          = "{The Neutrino Flux prediction at MiniBooNE}",
      journal        = "Physical Review D",
      year           = "2009",
      volume         = "79",
      page           = "072002"
      
      %eprint         = "https://doi.org/10.1103/PhysRevD.100.022001"
}

@article{majumdar:2021xse,
      author         = "K. Majumdar and K. Mavrokoridis",
      title          = "{Review of Liquid Argon Detector Technologies in the Neutrino Sector}",
      journal        = "Journal of Applied Sciences",
      year           = "2021",
      volume         = "11",
      issue          = "6",
      page           = "2455",
      eprint         = "https://doi.org/10.3390/app11062455"                                          
}

@article{abdulkhaleketal:2021lpe,
      author         = "R. Abdul Khalek \emph{et al.}",
      title          = "{Science Requirements and Detector Concepts for the Electron-Ion Collider:EIC Yellow Report}",
      year           = "2021",
      eprint         = "arXiv:2103.05419v3"                                          
}
\end{document}